\documentclass[article,nofootinbib,twocolumn, pra]{revtex4-2} 
\usepackage{amsmath,amssymb,multirow,epsfig,bm,pifont}
\usepackage[linkcolor=black,citecolor=black,urlcolor=blue,colorlinks=true,linktocpage=true]{hyperref}
\usepackage{graphicx}
\usepackage{epsfig}
\usepackage{bm}
\usepackage{tabularx}
\usepackage{makecell}

\newcommand{\beq}{\begin{equation}}
\newcommand{\eeq}{\end{equation}}
\newcommand{\bea}{\begin{eqnarray}}
\newcommand{\eea}{\end{eqnarray}}
\newcommand{\tr}{\mathrm{tr}}

\newcommand{\orcid}[1]{\href{https://orcid.org/#1}{\includegraphics[height=1.9ex,width=1.9ex]{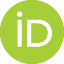}}}

\begin{document}

\title{An unconventional deformation of the nonrelativistic spin-1/2 Fermi gas}

\author{Vimal Palanivelrajan\ \orcid{0000-0001-9539-9218}}
\affiliation{Department of Physics and Astronomy, University of North Carolina, Chapel Hill, NC, 27599, USA}

\author{Joaqu\'in E. Drut\ \orcid{0000-0002-7412-7165}}
\affiliation{Department of Physics and Astronomy, University of North Carolina, Chapel Hill, NC, 27599, USA}

\begin{abstract}
We explore a generalization of nonrelativistic fermionic statistics that interpolates between bosons and fermions, in which up to 
$K$ particles may occupy a single-particle state. We show that it can be mapped exactly to $K$ flavors of fermions
with imaginary polarization. In particular, for $K\!=\!2$, we use such a mapping to derive the virial coefficients and relate them
to those of conventional spin-1/2 fermions in an exact fashion. We also use the mapping to derive next-to-leading-order perturbative results for 
the pressure equation of state. Our results indicate that the $K\!=\!2$ particles are more strongly coupled than conventional spin-$1/2$ 
fermions, as measured by the interaction effects on the virial expansion and on the pressure equation of state. 
In the regime set by the unitary
limit, the proposed $K\!=\!2$ deformation represents a universal many-body system whose properties remain largely unknown. In 
particular the system can be expected to become superfluid at a critical temperature $T_c$ higher than that of the unitary limit.
We suggest it may be possible to realize this system experimentally by engineering a polarized coupling to an electrostatic potential.
Finally, we show that the $K\!=\!2$ system does not display a sign problem for determinantal Monte Carlo calculations, which
indicates that $T_c$ can at least in principle be calculated with conventional methods.
\end{abstract}

\date{\today}

\maketitle

\section{Introduction}

Over the last two decades there has been considerable interest in the exploration
of universality in nonrelativistic quantum many-body systems (besides the well-known 
cases close to continuous phase transitions). By far the most studied case, both 
in theory and experiments (see e.g.~\cite{ZwergerBook, StrinatiReview, UltracoldAtoms4}), is that of 
spin-$1/2$ fermions in the unitary limit (a system of
nonrelativistic spin-$1/2$ particles with a zero-range interaction tuned
to the threshold of two-body bound-state formation, i.e. infinite scattering length). 
In such a situation, the property of universality stems from the lack of physical scales (and 
corresponding scale invariance) associated with the attractive interaction, as a system in this limit 
presents as many dimensionful parameters as a noninteracting gas (albeit also displaying strong 
pairing correlations and becoming superfluid at low enough temperature).
In practice, fermions at unitarity are realized to an excellent approximation in ultracold atom
experiments~\cite{UltracoldAtoms4} and to a lesser extent in the dilute neutron matter layer of neutron 
stars~\cite{StrinatiReview}.

Given the interest in these types of universal systems, it becomes a relevant question 
whether there are other related systems that are also universal. Examples of such cases
were proposed by Nishida and Son in Refs.~\cite{NishidaSon1D4B}, where they showed that there is a 
one-dimensional realization of the unitary limit with four flavors and a fine-tuned four-body 
interaction. They also showed, in a previous publication~\cite{UFGConformal}, that fermions at unitarity obey
a nonrelativistic conformal algebra for which anyons in two spatial dimensions provide a representation.

Motivated in part by the above developments, we explore here a definition of yet another type of particle statistics 
that presents nontrivial behavior at unitarity and in some sense interpolates between fermions
and bosons. Generally speaking, systems of quantum particles with unconventional statistics have 
been studied for many years (see e.g.~\cite{KhareBook}). The particular case we consider here is perhaps most directly related 
to the so-called Gentile statistics~\cite{Gentile}, which generalizes fermions and bosons by allowing 
single-particle states to hold at most $K$ particles, where $K\to 1$ recovers the fermionic case
and $K\to \infty$ the bosonic one (see also ~\cite{Green}).

As detailed below, our approach consists in starting with a fermionic system in a path-integral
representation and define the generalized $K$ statistics by modifying the integrand in a well-defined
fashion. We show that such a prescription corresponds to $K$ flavors of fermions with complex chemical potentials.
We then explore the thermodynamics of the system for the case $K=2$ at unitarity using
a perturbative approach as well as the virial expansion.

\section{Formalism}
\subsection{Noninteracting systems}

For completeness, we briefly review the noninteracting thermodynamics of fermions and bosons in the grand-canonical ensemble.
The grand canonical partition function is
\beq
\mathcal Z_0 = \tr \left[e^{-\beta (\hat T - \mu \hat N)} \right],
\eeq
where $\beta$ is the inverse temperature, $\mu$ the chemical potential, 
\beq
\hat N = \sum_{\bf p} \hat a^\dagger_{\bf p} \hat a_{\bf p},
\eeq
is the particle number operator, and
\beq
\hat T = \sum_{\bf p} \hat a^\dagger_{\bf p} \hat a_{\bf p} \epsilon({\bf p}),
\eeq
is the kinetic energy operator. Here, $\hat a^\dagger_{\bf p},  \hat a_{\bf p}$ are, respectively, the creation and annihilation operators for particles
of momentum ${\bf p}$,which satisfy the appropriate commutation or anticommutation relations.
We will set
\beq
\epsilon({\bf p}) = \frac{{\bf p}^2}{2m}
\eeq
at the end of the calculation, but as we show below, it is useful to keep $\epsilon({\bf p})$
as an arbitrary function.

For noninteracting fermions (one species),
\beq
\mathcal Z_{0,F} = \det[1 + z\mathcal U_0],
\eeq
where $\left[\mathcal U_0\right]_{{\bf p},{\bf p}'} = e^{-\beta \epsilon({\bf p})}\delta_{{\bf p},{\bf p}'}$ is a diagonal matrix 
in momentum space and $z = e^{\beta \mu}$ is the fugacity. 
Thus,
\beq
\ln \mathcal Z_{0,F} = \sum_{\bf p} \ln\left[1 + z e^{-\beta \epsilon({\bf p})}\right],
\eeq
which in the thermodynamic limit of large volume, yields a well-known integral expression that
is commonly written in terms of the so-called Fermi function~\cite{Huang}.

Similarly, for noninteracting bosons (again only one species),
\beq
\mathcal Z_{0,B} = \det[(1 - z\mathcal U_0)^{-1}],
\eeq
such that
\beq
\ln \mathcal Z_{0,B} = \sum_{\bf p}\ln\left[\left(1 - z e^{-\beta \epsilon({\bf p})}\right)^{-1}\right],
\eeq
which is also usually written in integral form in the thermodynamic limit.

Based on the above, it seems natural to consider defining a quantum statistics of identical particles that 
interpolates between the above two cases and is such that at most $K$ particles can occupy a given 
single-particle state. Such a statistics has been pursued by many authors in the past, perhaps the best known 
case being that of Gentile~\cite{Gentile}. For a maximum of $K$ particles per single-particle state, one obtains
\beq
\label{Eq:Z0K}
\mathcal Z_{0,K} = \det \left [\sum_{n=0}^{K} z^n \mathcal U_0^n \right],
\eeq
where the (single-flavor) fermionic case is recovered for $K=1$ and the bosonic case for $K\to \infty$.
As we show below, the $K=2$ is different from spin-$1/2$ fermions.
The polynomial in $x = z \mathcal U_0$ inside the determinant can naturally be written as
\beq
1 + x + x^2 + \dots + x^{K}= \frac{1 - x^{K+1}}{1-x},
\eeq
which is an easy way to see that the roots of our polynomial are $K$ of the $(K+1)$-th roots of unity, 
namely $\alpha_n = e^{i 2\pi n/(K+1)}$, with $n=1,2,\dots,K$, such that 
\beq
\ln \mathcal Z_{0,K} = \sum_{n=1}^{K} \sum_{\bf p} \ln \left(z e^{-\beta \epsilon({\bf p})} - \alpha_n\right),
\eeq
and therefore (dropping an additive constant which, in particular, vanishes if $K$ is even),
\beq
\ln \mathcal Z_{0,K} = \sum_{n=1}^{K} \sum_{\bf p} \ln \left(1 + w_n e^{-\beta \epsilon({\bf p})}\right),
\eeq
where $w_n = -\alpha_n z$ which shows that the noninteracting $K$ statistics corresponds to $K$ fermionic species with complex
fugacities.

As an example, consider the $K=2$ case, where
\beq
\label{Eq:K2Z0Det}
\mathcal Z_{0,2} = 
\prod_{\bf p} \left|1 + e^{-i\pi/3} z e^{-\beta \epsilon({\bf p})}\right |^2,
\eeq
such that total particle number is
\beq
N^{(0)} = z\frac{\partial \ln \mathcal Z_{0,2}}{\partial z} = 
\sum_{\bf p} 2 \text{Re} \left[\frac{e^{-i\pi/3} z e^{-\beta \epsilon({\bf p})}}{1 + e^{-i\pi/3} z e^{-\beta \epsilon({\bf p})}}\right],
\eeq
and such that the noninteracting occupation probabilities of the single-particle momentum states are
\beq
n_p^{(0)}(z) = 
2 \text{Re} \left[\frac{e^{-i\pi/3} z e^{-\beta \epsilon({\bf p})}}{1 + e^{-i\pi/3} z e^{-\beta \epsilon({\bf p})}}\right].
\eeq

\begin{figure}[h]
	\centering
	\includegraphics[width=\columnwidth]{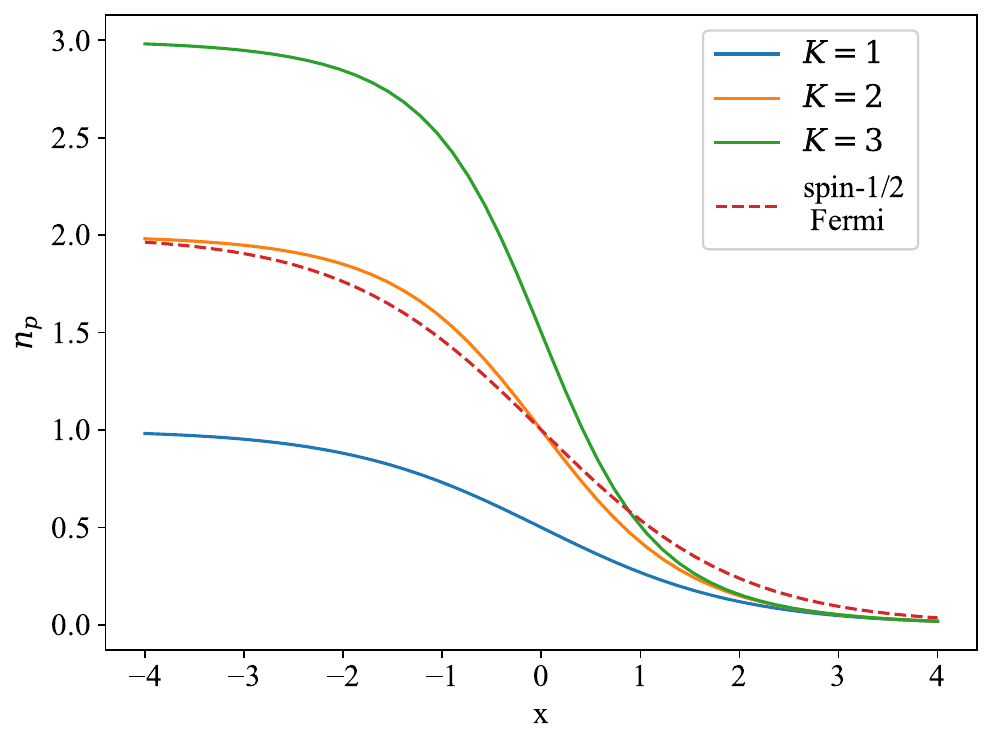}
	\caption{
		\label{fig:NoninteractingOccupation}
		Momentum distribution of the noninteracting $K=1,2,3$ gas.
		Here, the $x$ axis represents $x = \beta (p^2/(2m) - \mu)$.
		For comparison, we also show the noninteracting spin-$1/2$ Fermi gas result.
		}
\end{figure}

In Fig.~\ref{fig:NoninteractingOccupation} we show the momentum distribution $n_p^{(0)}(z)$ for
$K=2$, alongside the distributions for $K=1$ and $K=3$ and the spin-$1/2$ noninteracting Fermi gas.
Notably, there is a clear difference between the $K=2$ case and the spin-$1/2$ Fermi gas. The former
allows for higher occupations for $x<0$ at the cost of lower occupations at $x > 0$; both distributions
yield the same answer at $x=0$. 
As we show below when discussing the virial expansion, there is a well-defined sense in which
the $K=2$ case is "more bosonic" than the spin-$1/2$ Fermi gas, even though the former is effectively just 
a complex-$z$ deformation of the latter. 

%

\subsection{Interacting systems}

While the above completely determines the thermodynamics of the non-interacting case without further 
need to invoke the algebraic properties of $\hat a$ and $\hat a^\dagger$, the interacting case requires more care,
as interactions will involve products of four or more of these operators.

In this work we will follow a different, non-algebraic route by defining the interacting system using the field-integral formulation of the many-body problem. [We do, however, carry out our derivations using the spin-$1/2$ fermionic case as a starting point.] Using a Hubbard-Stratonovich transformation~\cite{Stratonovich,Hubbard} (see also~\cite{Hirsch,DrutNicholson}) to decouple the
interaction in the fermionic case, one obtains an expression for $\mathcal Z$ that involves a field
integral over auxiliary-field configurations $\sigma({\bf r},t)$ in which the integrand takes the form of a
product of two determinants:
\bea
{\det}\left(1 + z \mathcal U[\sigma]\right) {\det}\left(1 + z \mathcal U[\sigma]\right) \nonumber \\
= \det (1 + 2 z \mathcal U[\sigma] + z^2 \mathcal U&2[\sigma]),
\eea
which represent noninteracting systems immersed in the external field $\sigma({\bf r},t)$. We then deform such a system to our $K=2$ case by replacing the fermionic determinants as follows:
\bea
\label{Eq:K2ZDet}
&&{\det}\left(1 + z \mathcal U[\sigma]\right) {\det}\left(1 + z \mathcal U[\sigma]\right)\nonumber \\
&&\to \det (\alpha + z \mathcal U[\sigma])\det (\alpha^*+ z \mathcal U[\sigma]),
\eea
where $\alpha = -e^{i 2\pi/3} = e^{-i \pi/3}$ and where we have factored out the fugacity $z$ explicitly.

 The $\mathcal U[\sigma]$ matrices contain the product of $N_\tau$ exponentials of kinetic and potential energy operators along the imaginary-time direction:
\beq
\mathcal U[\sigma] = e^{-\tau T}e^{-\tau V[\sigma]}\dots e^{-\tau T}e^{-\tau V[\sigma]},
\eeq
where the inverse temperature is $\beta = N_\tau \tau$; $T$ is the single-particle representation of the kinetic energy operator;  and $V[\sigma]$ is the auxiliary potential resulting from the Hubbard-Stratonovich transformation.

We therefore define the interacting $K=2$ partition function as
\bea
\mathcal Z &=& \int \mathcal D \sigma \det (z \mathcal U[\sigma] + \alpha)\det (z \mathcal U[\sigma] + \alpha^*) \nonumber \\
&=& \int \mathcal D \sigma \det (1 + z \mathcal U[\sigma] +z^2 {\mathcal U}^2[\sigma]),
\eea
where we see that the expected $K=2$ noninteracting limit is recovered when the interaction is turned off.
Re-arranging, noting that $|\alpha|^2 = 1$, one obtains
\beq
\mathcal Z = \int \mathcal D \sigma \det (1 + w \mathcal U[\sigma])\det (1 + w^* \mathcal U[\sigma]),
\eeq
which shows once again that the $K=2$ case is identical to that of spin-$1/2$ fermions with complex fugacities $w=\alpha z$ and $w^*$ for each spin projection, respectively. In spite of the appearance of a complex fugacity, the presence of its complex conjugate indicates that this system does not display a sign problem for conventional
auxiliary-field Monte Carlo calculations, as long as the interaction is purely attractive.
Notably, $\alpha$ is temperature-independent, which leads to a temperature-varying complex effective chemical
potential, namely
\beq
\label{Eq:MuEff}
\beta \mu_\text{eff} = \ln w = \beta \mu \pm i \frac{\pi}{3},
\eeq
for spin-up and spin-down, respectively. Thus, our $K=2$ system is equivalent to a spin-$1/2$ Fermi gas with 
imaginary polarization.

It is well-known~\cite{Hasenfratz} that a chemical potential is equivalent to an imaginary $A_0$ gauge field.
Thus, coupling to a constant $A_0$ gauge field is equivalent to an imaginary chemical potential, which is our interest here.
Therefore, the above imaginary polarization can potentially be realized by engineering a coupling of spin-$1/2$ fermions
in which the spins couple to an electrostatic field with an electrostatic potential difference of $2\pi/3$.
Ideally, this coupling would be independent from the usual Feshbach resonance coupling that controls the inter-spin interaction.

In the next sections we will focus on this $K=2$ case, tuning the interaction to the unitary limit. While it is generally accepted that 
polarized unitary fermions undergo a phase transition at some polarization, from a superfluid phase to a normal phase (possibly 
going through exotic superfluid phases), their behavior at imaginary polarization remains little explored (see however the seminal 
studies of Refs.~\cite{ImaginaryPolarizationPRL, ImaginaryPolarization1DPRA}).

In investigating the present deformation of the Fermi gas, we have kept the internal number of degrees of 
freedom constant, i.e. we compare $K=2$ to the spin-$1/2$ Fermi gas. Another route could be to 
promote the individual spin degrees of freedom of the Fermi gas to $K=2$, but this would result in a
system with four internal degrees of freedom, which would be generally very difficult to compare with
its spin-$1/2$ counterpart and possibly unstable depending on the form of the interaction (see below).

For $K > 2$, attractive interactions make the unitary-limit system unstable toward Thomas collapse (due to the formation of infinitely
deep three-body bound states)~\cite{ThomasCollapse} (see also~\cite{RMPClusters}). However, one may still envision
an appropriate modification of the interaction at short range along with a decoupling of the interaction to proceed with 
the mapping to fermions by factorization of the determinant as
\beq
\det \left [\sum_{n=0}^{K} z^n \mathcal U^n[\sigma] \right] = \prod_{n=1}^{K}\det \left[1 + w_n \mathcal U[\sigma]\right],
\eeq
where $w_n = \alpha_n z$, and $\alpha_n$, $n=1,2,\dots,K-1$ is one of the (non-unity) $(K+1)$-th complex roots of unity.
In other words, our system maps exactly onto a system of $K$ fermions with complex chemical potentials $w_n$.
For $K$ even, the relevant roots of unity come in complex conjugate pairs, such that pairing the determinants accordingly
one finds that there is no sign problem. For $K$ odd, on the other hand, there will always be a determinant factor of the form
\beq
\det (1 - z\mathcal U[\sigma]),
\eeq
which can be made real (in some cases, depending on the form of the interaction), but cannot be guaranteed in general to be 
positive definite (except under specific symmetry conditions; see e.g. Ref.~\cite{SignProblemReview} for a review); such a factor corresponds to the root 
$\alpha=-1$, which is present for all $K$ odd.

The next few sections present our results for $K=2$ for the virial coefficients up to fourth order, and a next-to-leading-order perturbative
calculation of the pressure.

\section{Results: The virial expansion}

The virial expansion (see e.g.~\cite{Huang}) is an expansion of $\mathcal Z$ in powers of the fugacity $z$, such that
\beq
\mathcal Z = \sum_{n=0}^{\infty} Q_n z^n,
\eeq
where $Q_n$ is the $n$-particle canonical partition function, and
\beq
\ln \mathcal Z = Q_1\sum_{n=1}^{\infty} b_n z^n,
\eeq
where $b_n$ are the virial coefficients, typically written in terms of $Q_m$ with $m \leq n$;
for example, $b_1 = 1$, while
\beq
b_2 = \frac{Q_2}{Q_1} - \frac{Q_1}{2!},
\eeq
and
\beq
b_3 = \frac{Q_3}{Q1} + b_2 Q_1- \frac{Q_1^2}{3!},
\eeq
and so forth, where $Q_1 = V/\lambda_T^3$ in 3D.
The above expressions for $b_2$ and $b_3$ are independent of the quantum statistics.

For reference, we note that the $b_n$ for noninteracting fermions and bosons 
in homogeneous space are given, respectively, by
\beq
b_{0,F,n} = (-1)^{n+1}\frac{1}{n^{5/2}},
\eeq
and
\beq
b_{0,B,n} = \frac{1}{n^{5/2}}.
\eeq

It is not difficult to calculate the virial coefficients of the noninteracting case
for arbitrary $K$ and notice that they are identical to those of the bosonic case,
except that for every $n$ that is multiple of $K+1$ one obtains an extra sign and an overall
factor of $K$, i.e. 
\beq
b_{0,K, m(K+1)} = - K \frac{1}{n^{5/2}},
\eeq
where $m = 1,2,3\dots$, which also captures the expected results both at $K=1$ and $K \to \infty$.

It is straightforward to derive relations between the virial coefficients of the $K=2$ gas
and the spin-$1/2$ Fermi gas by noting that the virial expansion 
for the complex-polarized Fermi gas takes the form
\bea
\ln (\mathcal Z/\mathcal Z_0) = 2 Q_{1,0}^F \sum_{n=2}^{\infty} \sum_{m + j = n} \Delta b^F_{m,j} z_{\uparrow}^{m}z_{\downarrow}^{j},
\eea
where $Q_{1,0}^F  = V/\lambda_T^3 = Q_1$, $z_{\uparrow} = z e^{i \alpha}$, and $z_{\downarrow} = z e^{-i \alpha}$, and $\Delta b^F_{m,j}$ are the virial coefficients of the polarized spin-$1/2$ Fermi gas.
Thus, identifying the powers of $z$ against the $K=2$ expression 
\beq
\ln (\mathcal Z/\mathcal Z_0) = Q_{1,0}^F \sum_{n=2}^{\infty} \Delta b^{K=2}_{n} z^{n} ,
\eeq
we find
\beq
\Delta b^{K=2}_2 = 2 \Delta b^F_{11} = 2 \Delta b^F_2,
\eeq
\beq
\Delta b^{K=2}_3 = 2 \Delta b^F_{21} = \Delta b^F_3,
\eeq
\beq
\Delta b^{K=2}_4 = - 2 \Delta b^F_{31} + 2 \Delta b^F_{22} \neq \Delta b^F_{4},
\eeq
where for completeness we note that $\Delta b^{F}_4 = 2 \Delta b^F_{31} + \Delta b^F_{22}$.

Using known results at unitarity from Refs.~\cite{VE1, VE2, VE3}, we find
\beq
\Delta b^{K=2}_2 = \sqrt{2},
\eeq
\beq
\Delta b^{K=2}_3 = -0.3551...,
\eeq
\beq
\Delta b^{K=2}_4 = -0.435...,
\eeq
where we only quote enough digits for the purposes of this work.
It is evident from the above, however, that for $K=2$ at unitarity, the fourth-order coefficient
is larger in magnitude than the third-order coefficient. This is in contrast to the conventional
spin-$1/2$ unitary Fermi gas, where that type of behavior shows up at one higher order (i.e.
between fourth and fifth).

For future reference, we note that the second-order virial expansion of the pressure 
at unitarity reads
\beq
\beta \Delta P V = \ln (\mathcal Z/\mathcal Z_0) = Q_{1,0}^F  \sqrt{2} z^{2},
\eeq 
such that, using the noninteracting result $\beta P_0 V = Q_1 (z + b_2^{0}z^2 + \dots)$,
we obtain
\beq
\frac{P}{P_0} = 1 + \frac{\Delta P}{P_0} = 1 + \frac{\sqrt{2} z^{2}}{z + b_2^{0}z^2 + \dots} = 1+\Delta b^{K=2}_2 z
\eeq

Carrying out the expansion now up to $z^3$ order, we obtain
\bea
\frac{P}{P_0} &=& 1+ \Delta b^{K=2}_2 z + \left [-\frac{\Delta b^{K=2}_2 }{4\sqrt{2}}+\Delta b^{K=2}_3 \right] z^2  \nonumber \\
&&\!\!\!\!\!+ \left[-\frac{\Delta b^{K=2}_3}{4 \sqrt{2}}+\Delta b^{K=2}_4+\Delta b^{K=2}_2 \left(\frac{1}{32}+\frac{2}{9\sqrt{3}}\right) \right ] z^3. \nonumber
\eea
%

\section{Results: Perturbative approach}

To complement our VE results of the previous section, we present here a
second-order perturbation theory calculation of the thermodynamics of our $K=2$ 
system for arbitrary $w$. To this end, our starting point is
\beq
\mathcal Z = \int \mathcal D \sigma \det (1 + w \mathcal U[\sigma])\det (1 + w^* \mathcal U[\sigma]).
\eeq
To expand $\mathcal Z$ perturbatively, we use a discretization of the imaginary-time 
direction, as in Ref.~\cite{Morrell}, and expand $\mathcal U[\sigma]$ in powers of the
bare coupling constant $C$, such that
\beq
\mathcal U[\sigma] = \mathcal U_0 + C \mathcal U_1[\sigma]+ C^2 \mathcal U_2[\sigma] + \dots,
\eeq
where, we define $C$ by
\beq
e^{-\tau V[\sigma]} = 1 + C M[\sigma],
\eeq
and $M[\sigma]$ contains all the non-trivial dependence on $\sigma$ and its form
will depend on the specific choice of Hubbard-Stratonovich transformation.
Then
\bea
\det (1 + w \mathcal U[\sigma]) 
&=& \det (1 + w \mathcal U_0) \\
&& \times \det \left [1 + C  X_1[\sigma]+ C^2  X_2[\sigma] \right], \nonumber \\
\eea
where
\beq
X_k[\sigma] = \frac{w\mathcal U_k[\sigma]}{1 + w \mathcal U_0}.
\eeq

Therefore, at order $C^2$,
\bea
\det (1 + w \mathcal U[\sigma]) 
&=&
1 + C \tr X_1[\sigma]\nonumber + C^2 \tr X_2[\sigma] \\
&&+ 
\frac{C^2}{2}  
\left[\tr^2 X_1[\sigma] - \tr X^2_1[\sigma]\right].
\eea

Integrating over $\sigma$ to get $\mathcal Z$, we obtain
\bea
\label{Eq:DetProductwwstar}
\mathcal Z &=& \int \mathcal D \sigma \det (1 + w \mathcal U[\sigma])\det (1 + w^* \mathcal U[\sigma]) \nonumber \\
&=& \mathcal Z_0 \left[1 + C \Delta_1(w,w^*) + C^2 \Delta_2(w,w^*) \right],
\eea
where the noninteracting partition function is
\beq
\mathcal Z_0 = \det (1 + w \mathcal U_0)\det (1 + w^* \mathcal U_0),
\eeq
the first-order contribution in $C$ is
\beq
\Delta_1(w,w^*) = \int \mathcal D \sigma \left( \tr X_1[\sigma] +  c.c. \right)
\eeq
and the second-order term is
\bea
\Delta_2(w,w^*) &=& \int \mathcal D \sigma [ \tr X_1[\sigma] \tr X^*_1[\sigma] + \tr X_2[\sigma] \nonumber \\
&&+  \frac{1}{2} \left (\tr^2 X_1[\sigma] - \tr X^2_1[\sigma] + c.c. \right )  ].
\eea

Since we assume the interaction to be only a two-body interaction, only terms with
even powers of $w$ will contribute to the final result. In the above equations, it can be
shown that, indeed, $\Delta_1 = 0$. Similarly, the $X_2$ term in $\Delta_2$ also
vanishes. The remaining contribution in $\Delta_2$ is what is conventionally
called the first-order perturbation theory result, which boils down to
\bea
\mathcal Z / \mathcal Z_0 = 1 + C^2 V n(w) n(w^*),
\eea
where
\beq
n(w) = \frac{1}{V}\sum_{p} n_p(w),
\eeq
and
\beq
n_p(w) = \frac{w e^{-\beta \epsilon(p)}}{1 + w e^{-\beta \epsilon(p)}}.
\eeq
In the large-volume limit,
\beq
n(w) \to \frac{V}{(2\pi)^3} \int dp\, n_p(w),
\eeq
and setting $\epsilon(p) = p^2/(2m)$ the integral can be evaluated and is proportional
to $Li_{\frac{3}{2}}(-w)$.

Thus, the change in pressure is given at this order by
\beq
\beta \Delta P V =  \ln \left( \mathcal Z / \mathcal Z_0 \right) = C^2 V n(w) n(w^*),
\eeq
such that our final result, setting $\epsilon(p) = p^2/(2m)$, is 
\beq
\label{Eq:PoP0}
\frac{P}{P_0} = 1 + \mathcal{C}\left| Li_{\frac{3}{2}}(-w) \right|^2 [2\text{Re}Li_{\frac{5}{2}}(-w)]^{-1},
\eeq
where $\mathcal{C}$ is a dimensionless coupling to be renormalized as explained below.
Note that we have used the noninteracting result
\bea
\beta P_0 V =  \ln \mathcal Z_0 = 2\text{Re} \sum_p \ln \left[1 + w e^{-\beta \epsilon(p)} \right],
\eea
where the last sum, in the large-volume limit, is proportional to $Li_{\frac{5}{2}}(-w)$.

To renormalize $\mathcal C$, we use the VE result of the previous section by choosing a renormalization point $z_0$.
For our $K = 2$ case, with $\alpha = e^{-i\pi/3}$, the second-order VE for the pressure reads
\beq
\frac{P}{P_0} = 1 + \sqrt{2}z,
\eeq
and therefore
\beq
\mathcal{C} = \sqrt{2}z_0\left| Li_{\frac{3}{2}}(-e^{-i\pi/3}z_0) \right|^{-2}[2 \text{Re}Li_{\frac{5}{2}}(-e^{-i\pi/3}z_0)],
\eeq
which fixes $\mathcal C$ in Eq.~(\ref{Eq:PoP0}).

\begin{figure}[h]
	\centering
	\includegraphics[width=\columnwidth]{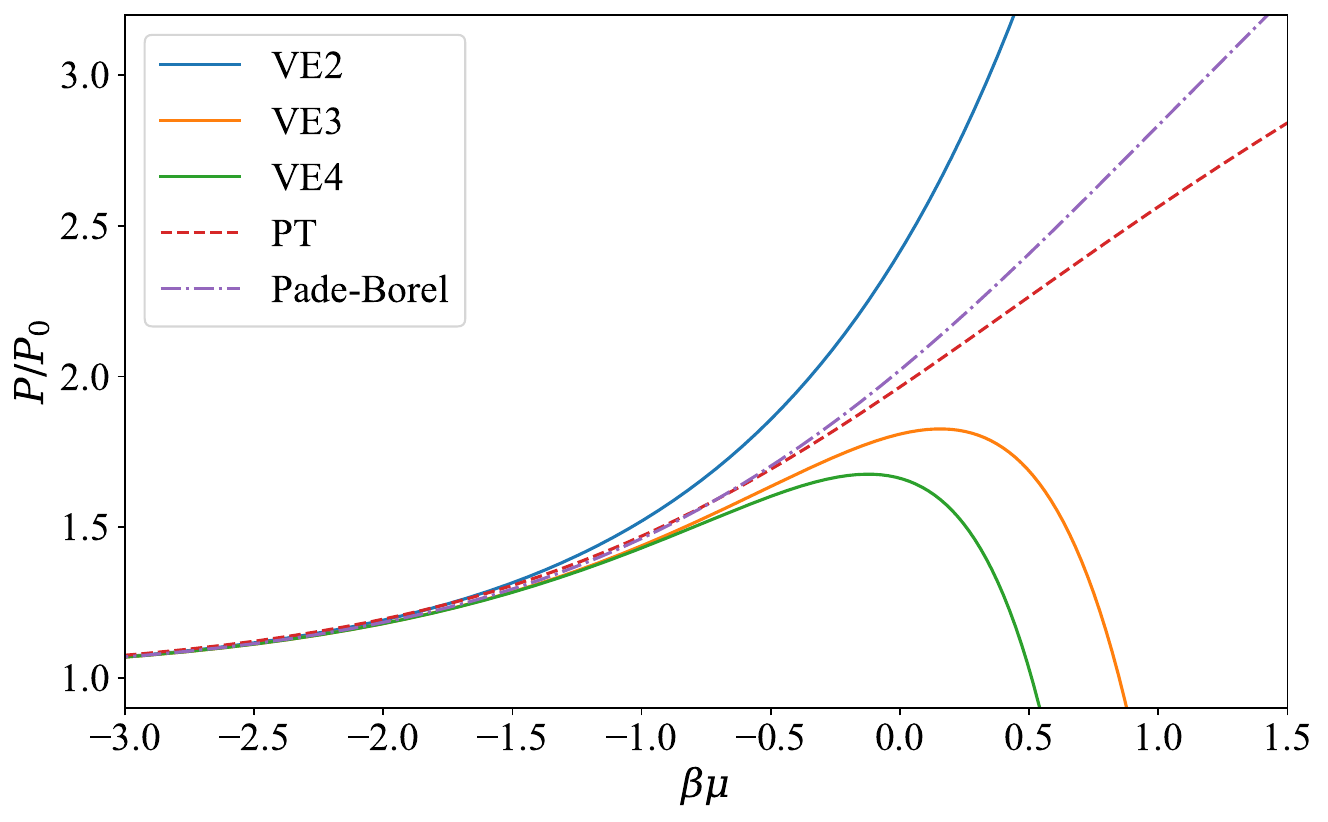}
	\caption{
		\label{fig:Pressure}
		Pressure $P$ of the $K=2$ gas at unitarity, in units of its noninteracting counterpart $P_0$ as a result of four different calculations: virial expansion at second, third, and fourth orders, and first-order perturbative result.
We also show the result of a Pade-Borel resummation of the fourth-order virial expansion.
		}
\end{figure}

In Fig.~\ref{fig:Pressure} we show our results for the pressure $P$ in units of $P_0$, 
comparing the second, third, and fourth-order virial expansions with our perturbative
result for $P/P_0$ as a function of $\beta \mu$, at unitarity for $K=2$.

%

\begin{figure}[t]
	\centering
	\includegraphics[width=\columnwidth]{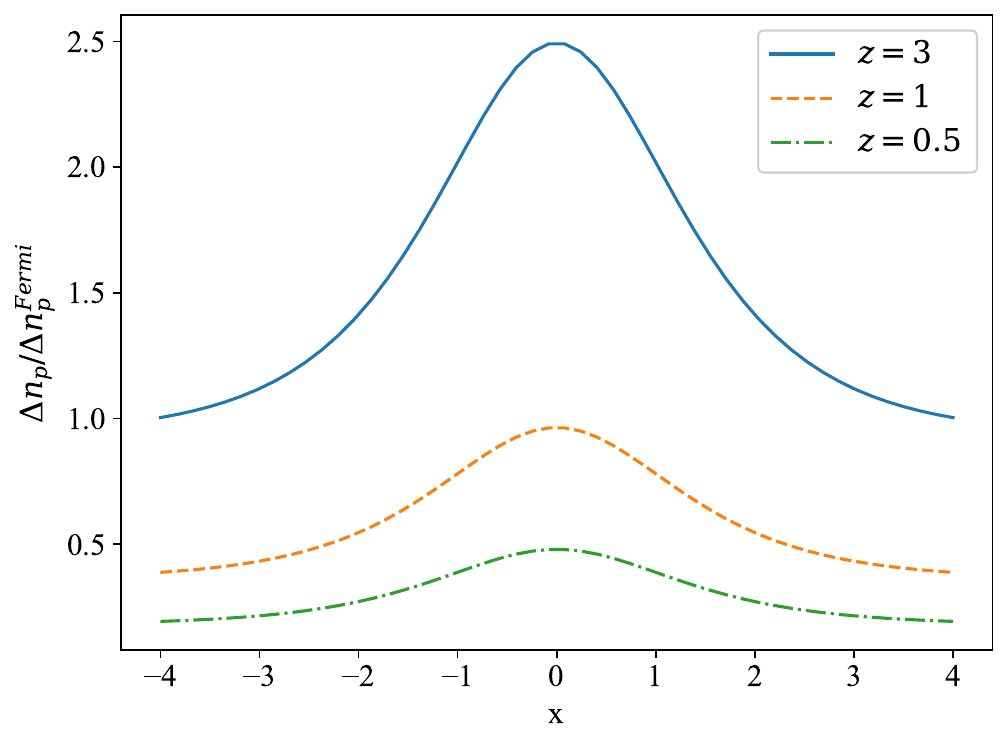}
	\caption{
		\label{fig:InteractingOccupationCorrection}
		Momentum distribution first-order perturbative correction for $K=2$,
		relative to the conventional spin-$1/2$ Fermi gas,.
		Here, the $x$ axis represents $x = \beta (p^2/(2m) - \mu)$.
		For comparison, results are shown for various fugacities $z=0.25$, $1.0$, and
		$3.0$.
		}
\end{figure}

Using the above expressions, we may also access the momentum distribution.
For that purpose, we restore a generic dispersion relation $\epsilon(p)$ instead of
$p^2/(2m)$, which allows us to reinterpret the expressions for $\mathcal Z$ as
generating functionals for expectation values of the occupation probability $n_p$.
For instance, in the noninteracting limit,
\beq
n^{(0)}_p = -\frac{1}{\beta} \frac{\delta \ln \mathcal Z_0}{\delta \epsilon(p)} = 2 \text{Re}[n_p(w)].
\eeq
The interaction effects on the above are given by 
\bea
\Delta n^{}_p &=& -\frac{1}{\beta} \frac{\delta \ln (\mathcal Z/\mathcal Z_0)}{\delta \epsilon(p)} \nonumber \\
&\propto&\text{Re} \left[ n(w) \frac{(w^*)^{-1}e^{\beta \epsilon(p)}}{(1+(w^*)^{-1}e^{\beta \epsilon(p)})^2} \right].
\eea
In Fig.~\ref{fig:InteractingOccupationCorrection}, we show the above correction $\Delta n^{}_p$
relative to its conventional spin-$1/2$ counterpart $\Delta n^\text{Fermi}_p$, for three different fugacities,
as a function of $x = \beta (p^2/(2m) - \mu)$. 

At sufficiently low- and high-$x$, $\Delta n^\text{Fermi}_p$ tends to zero.
Thus, we note that $\Delta n^{}_p$ also decreases to zero at sufficiently low- and high-$x$. This indicates that
the maximum occupation number (i.e. 2 for $K=2$) is not modified by the interactions, at least away from $x=0$.
Around $x=0$, interaction effects may violate the maximum occupation number for sufficiently strong interactions
or sufficiently high $z$, but those effects are beyond our perturbative analysis.

At high fugacities, interaction effects are substantially enhanced for $K=2$ relative to the Fermi gas,
which is important because at sufficiently high $z$, these systems are expected to become superfluid. 
Based on our results for $\Delta n^{}_p$, we anticipate that the critical temperature for $K=2$ will be higher 
than that of the spin-1/2 Fermi gas at the same interaction strength, as the interparticle attraction is effectively 
stronger for the $K=2$ gas.

%

\section{Summary and conclusions}

In this work, we have explored a generalization of nonrelativistic fermionic statistics that interpolates between bosons and fermions, 
for which up to $K$ particles can occupy a single-particle state. We have shown that it can be mapped exactly to $K$ flavors of 
fermions with a specific temperature-dependent imaginary polarization, i.e. the difference in density among the flavors is an 
imaginary quantity. In particular, for $K=2$, 
we use the mapping to derive the virial coefficients and relate them to those of conventional spin-1/2 fermions in an exact fashion. 
We also use the mapping to derive next-to-leading-order perturbative results for the pressure equation of state. Our results indicate 
that the $K=2$ particles are more strongly coupled than conventional spin-$1/2$ fermions, as measured by the interaction effects 
on the virial expansion and the pressure. 

At unitarity, the proposed $K=2$ system is a universal many-body system whose properties remain largely unknown. In 
particular the system can be expected to become superfluid at a critical temperature $T_c$ higher than that of the conventional, 
unpolarized unitary limit~\cite{Tc1,Tc2,Tc3,Tc4}. Indeed, the mean-field study of Ref.~\cite{ImaginaryPolarizationPRL} found that, for an imaginary polarization of $\pi/3$ 
[see Eq.~(\ref{Eq:MuEff})], the critical temperature is about $30\%$ higher than in the unpolarized case. If that percent change
applies once all fluctuations are accounted for, one may expect $T_c/\epsilon_F \simeq 0.2$ for the $K=2$ system 
(where $\epsilon_F$ is the Fermi energy). By engineering a polarized coupling to an electrostatic potential, it may be possible to realize this system in a controlled fashion via ultracold atoms, where the question of $T_c$ can be explored experimentally.
Finally, we showed that the $K=2$ system does not display a sign problem for determinantal Monte Carlo calculations, which indicates that the precise value of $T_c$ can at least in principle be determined with conventional methods.

\acknowledgments
This research was funded by National Science Foundation Grant No. PHY2013078.



\end{document}